\documentclass[prb,onecolumn,floatfix]{revtex4}

\usepackage{amsfonts}
\usepackage{amsmath}
\usepackage{verbatim}
\usepackage[dvips]{graphicx}
\usepackage{subfigure}
\usepackage{sidecap}

\def\Tr{\textrm}
\def\dd{\textrm{d}}
\def\Bf{\boldsymbol}

\def\rv{\Bf{r}}

\def\kv{\Bf{k}}

\def\qv{\Bf{q}}
\def\Qv{\Bf{Q}}
\def\Gv{\Bf{G}}
\def\nv{\Bf{n}}

\begin{document}

\title{Pair density wave instability and Cooper pair insulators in gapped fermion systems}

\author{P. Nikoli\'c$^{1,2}$, A. A. Burkov$^{3}$ and A. Paramekanti$^{4}$}

\address{$^1$ Department of Physics and Astronomy, George Mason University, Fairfax, VA 22030, USA}
\address{$^2$ Department of Physics and Astronomy, Johns Hopkins University, Baltimore, MD 21218, USA}
\address{$^3$ Department of Physics and Astronomy, University of Waterloo, Waterloo,  Ontario N2L 3G1, Canada}
\address{$^4$ Department of Physics, University of Toronto, Toronto, Ontario M5S 1A7, Canada}



\begin{abstract}
By analyzing simple models of fermions in lattice potentials we argue that the zero-temperature pairing instability of any ideal band-insulator occurs at a finite momentum. The resulting supersolid state is known as "pair density wave". The pairing momentum at the onset of instability is generally incommensurate as a result of phase-space restrictions and relative strengths of interband and intraband pairing. However, commensurate pairing occurs in the strong-coupling limit and becomes a Cooper-channel analogue of the Halperin-Rice exciton condensation instability in indirect bandgap semiconductors. The exceptional sensitivity of incommensurate pairing to quantum fluctuations can lead to a strongly-correlated insulating regime and a non-BCS transition, even in the case of weak coupling as shown by an exact renormalization group analysis.
\end{abstract}

\maketitle

\section{Introduction}

One of the central challenges in the theory of high temperature superconductors is how to reconcile the mean-field character of Cooper pairing which is pivotal in any fermionic superfluid, and the seemingly fluctuation-driven superconducting transition in underdoped cuprates \cite{emery95}. Motivated by this problem, we consider $s$-wave pairing instability in generic models of band-insulators and argue that certain phenomena familiar from cuprates, such as density-wave patterns \cite{Davis2004b} and fluctuation-driven transitions, are possible in these simple systems. The most direct physical realization of our models can be found in ultra-cold gases of alkali atoms tuned near a broad Feshbach resonance \cite{Chin2006}.

The simplest generic band-insulator is obtained at zero temperature by placing an even number of fermionic particles in each unit-cell of a periodic potential $V_{\rv}$:
\begin{equation}\label{model1}
H = \int \dd^3 r \Biggl\lbrack c_{\sigma}^{\dagger}
  \left( -\frac{\boldsymbol{\nabla}^2}{2m}-\mu+V_{\rv} \right) c_{\sigma}^{\phantom{\dagger}}
  - U c_{\uparrow}^{\dagger} c_{\downarrow}^{\dagger}
         c_{\downarrow}^{\phantom{\dagger}} c_{\uparrow}^{\phantom{\dagger}}
   \Biggr\rbrack \ .
\end{equation}
We require that the chemical potential $\mu$ sit in a band-gap, but allow changes of $\mu$, lattice amplitude $V$ and attractive interaction strength $U$ to drive phase transitions (we use $\hbar=1$). In order to stay focused on universal aspects of unconventional pairing, we choose to work in the unitarity limit which defines a strongly-interacting quantum critical point at which the only relevant interaction, included in (\ref{model1}), is related to the two-body vacuum scattering length $a$ through a cut-off scale implicit in the following integral:
\begin{equation}\label{scatlength}
\frac{1}{U} = - \frac{m}{4\pi a} + \sum_{\nv}\int\frac{\dd^3 k}{(2\pi)^3}
  \frac{1}{2 \epsilon_{\nv\kv}}\Bigl\vert_{V=0} \ .
\end{equation}
Here $\nv$ and $\kv$ are band-index and crystal momentum quantum numbers of quasiparticles, and $\epsilon_{\nv\kv}$ are their bare energies.

\section{Pair density wave instability}

In our recent work \cite{Nikolic2009a} we argued that the pairing instability in the continuum model (\ref{model1}) generally occurs at a \emph{finite incommensurate} wavevector $\Qv$. Before subjecting this conclusion to more scrutiny, we outline the robust reasons for finite-momentum pairing. The mean-field inverse pairing susceptibility matrix $\Pi$ at frequency $\Omega=0$ in the insulating state is:
\begin{equation}\label{Bubble}
\Pi_{\Gv\qv;\Gv'\qv'} =
    \sum_{\nv_1\nv_2} \int \frac{\dd^3 k_1}{(2\pi)^3} \frac{\dd^3 k_2}{(2\pi)^3} \;
    \frac{f\left(\xi_{\nv_1 \kv_1}\right) - f\left(-\xi_{\nv_2 \kv_2}\right)}
         {\xi_{\nv_1 \kv_1} + \xi_{\nv_2 \kv_2}}
\Gamma_{\nv_1 \kv_1 ; \nv_2 \kv_2}^{\Gv \qv *} \Gamma_{\nv_1 \kv_1 ; \nv_2 \kv_2}^{\Gv' \qv'} +
             \frac{(2\pi)^3}{U} \delta(\qv-\qv')\delta_{\Gv \Gv'}
\end{equation}
where $\Gv$ are reciprocal lattice vectors, $\xi_{\nv \kv}= \epsilon_{\nv \kv} - \mu$, and $f(\xi)$ is Fermi-Dirac distribution function. The conservation of crystal momentum implies $\Pi_{\Gv\qv;\Gv'\qv'} = \Pi_{\Gv\Gv'}(\qv)\times (2\pi)^3 \delta(\qv-\qv')$. This expression is obtained by integrating out the internal Matsubara frequency in the bubble Feynman diagram, where the fermion propagators are written in the basis of Bloch states \cite{Nikolic2009a}. Such a basis choice introduces non-trivial momentum and band dependent vertex functions:
\begin{equation}\label{Ver}
\Gamma_{\nv_1 \kv_1 ; \nv_2 \kv_2}^{\Gv \qv} = \int\dd^3 r \; e^{-i(\qv+\Gv)\rv}
  \psi_{\nv_1 \kv_1}^{\phantom{*}}(\rv) \psi_{\nv_2 \kv_2}^{\phantom{*}}(\rv) \propto \delta(\kv_1+\kv_2-\qv) \ .
\end{equation}

\begin{figure}
\subfigure[{}]{\includegraphics[height=2.0in]{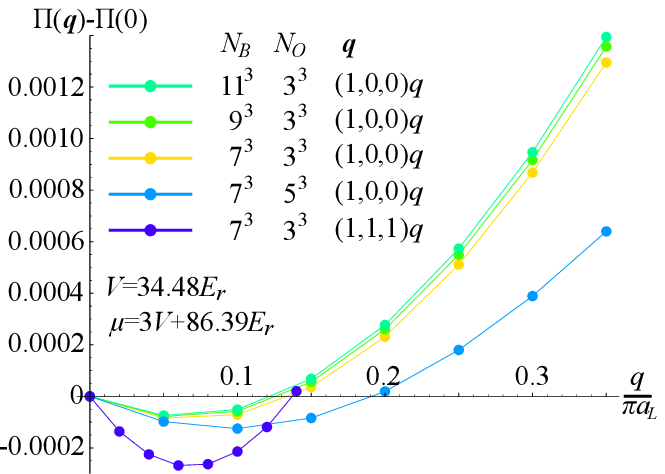}} \hspace{0.7in}
\subfigure[{}]{\includegraphics[height=2.0in]{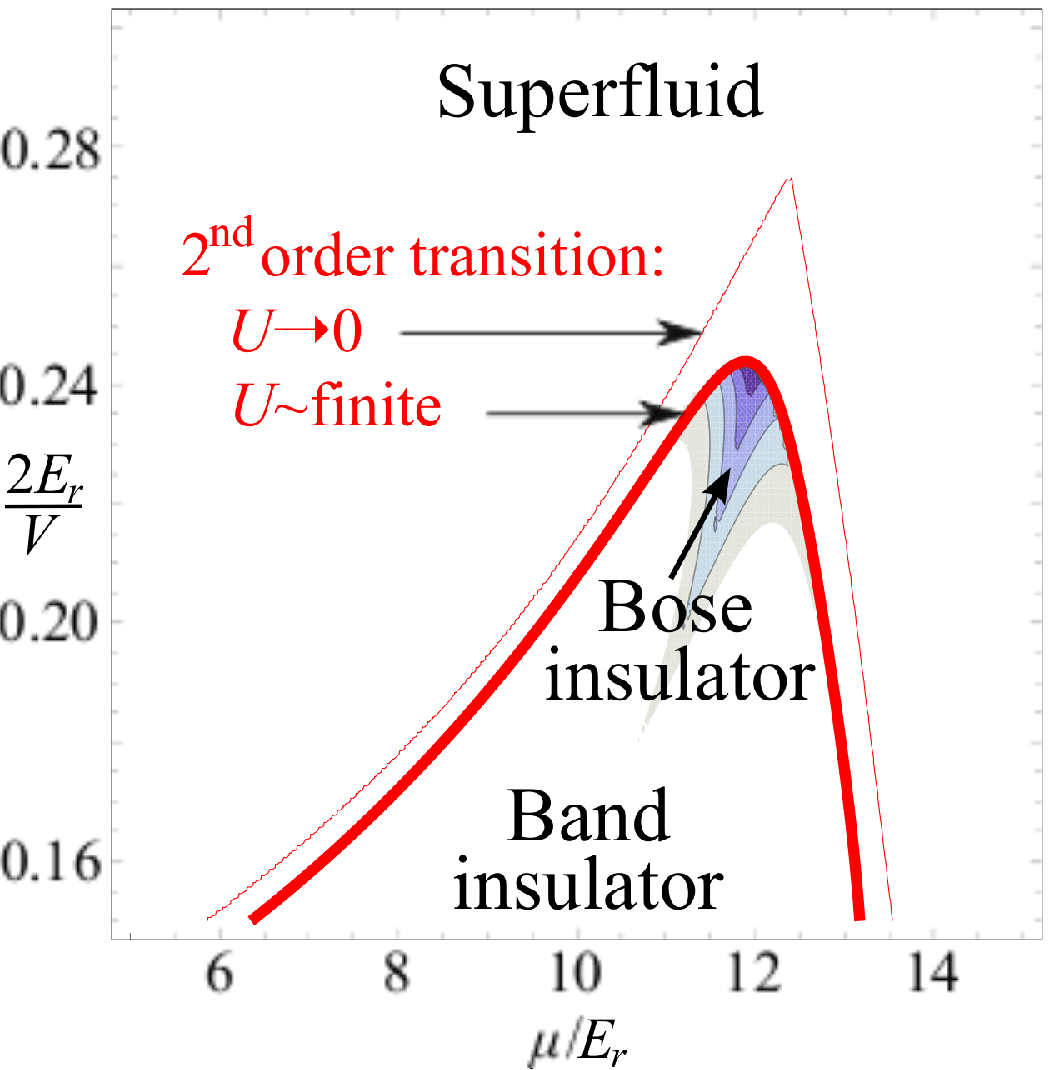}}
\caption{\label{PiPlot}(a) $\Pi(\qv)$ in $d=3$ for the potential $V_{\rv} = V\!\! \sum\limits_{i=1}^{d} \! \cos(2\pi x_i/a_L)$; $E_r \!\! = \!\! \pi^2\hbar^2/2ma_L^2$, $N_B$ is the number of bands, $N_O$ is the number of order parameters. (b) A $d=2$ phase diagram at $T=0$.}
\end{figure}

Typical plots of the lowest eigenvalue $\Pi(\qv)$ of (\ref{Bubble}) are shown in Fig.\ref{PiPlot}(a). These numerical calculations differ by cutoffs $N_B$ (the number of bands) and $N_O$ (the number of order parameter plane-wave harmonics at different $\Gv$), and also illustrate the dependence on the orientation of $\qv$. By keeping the density at two fermions per site we invariably find that $\Pi(q)$ is linear for small $q$ and reaches its minimum at a finite value of $q=|\Qv|$ in the direction $\hat{\qv}=(1,1,1)/\sqrt{3}$ for the simple cubic cosine potential. It can be seen from (\ref{Bubble}) that by varying the interaction strength $U$ all eigenvalues of the matrix $\Pi$ are shifted by the same amount. Pairing occurs when the smallest eigenvalue $\Pi(\qv)$ becomes negative, but this happens at a finite momentum $\Qv$.

The momentum dependence of $\Pi(\qv)$ enters through both the denominator in (\ref{Bubble}) and the vertex functions. The former can lead to \emph{commensurate} pairing instability if the interband pairing channels are strong enough, a particle-particle analogue of Halperin-Rice exciton condensation in indirect gap semiconductors \cite{Halperin1968}. The vertex functions, however, contain linear terms in $\qv$ which respect lattice symmetries (such as $\qv \nv_1$, etc.), arising from phase-space restrictions for intraband and interband pairing. Their contributions are not canceled out in (\ref{Bubble}) thanks to the non-linear factor involving $\xi_{\nv\kv}$, and make $\Pi(\qv) \sim A_{\hat{\qv}}|\qv|$ for small $q$ leading to an \emph{incommensurate} pairing instability. A negative value $A_{\hat{\qv}}<0$ is required in order to secure a local minimum at finite $\Qv$. We never found a positive $A_{\hat{\qv}}$ in our calculations, which would imply a uniform pairing instability with an unphysical dispersion for Goldstone modes.

Fluctuation corrections can be systematically included by promoting the theory to a simplectic Sp($2N$) symmetry group and performing a perturbative $1/N$ expansion \cite{nikolic:033608}. The corrections still involve the same vertex functions, so the linear-$q$ features are unlikely to be accidentally canceled. Therefore, we can expect incommensurate finite-momentum pairing instability to all orders of $1/N$.

However, an equilibrium superfluid phase with incommensurate pairing is not necessarily stable. The equilibrium order parameter must be a standing wave involving all symmetry-related wavevectors $\Qv$ at which pairing occurs most readily. The obtained state is called a pair-density-wave (PDW) and breaks translational symmetry both with its global U(1) phase and density. An incommensurate density modulation is very frustrated in a lattice potential and thus not pinned to the lattice. Quantum fluctuations can remove this frustration by stabilizing some commensurate order, at least at $\Qv=0$. The arguments above suggest that the linear-$q$ dependence of $\Pi(q)$ can hardly be completely removed, but fluctuations can modify $\Pi(q)$ to push its quadratically-dispersing minimum to a finite commensurate wavevector.

\section{Fluctuation effects}

The emergence of a fluctuation-dominated regime can be clearly seen in renormalization group (RG) analysis \cite{Nikolic2010}. The theory (\ref{model1}) contains a number of interacting fixed points associated with a diverging ``vacuum'' scattering length $a$ at zero temperature. Any band-insulator can be viewed as a ``vacuum'' of particles or holes and taken to the unitarity limit by adjusting the chemical potential to its nearest band-edge and tuning $U$ to a critical value $U^*$ appropriate for the given lattice potential. The universal theory near unitarity generally contains a single fermion species, either particles or holes, whose low-energy dynamics is entirely captured by an effective mass; fermionic excitations in other bands are gapped and thus irrelevant near the fixed point:
\begin{equation}\label{CritTheory1}
S' = \int \mathcal{D}k \; f_{k,\alpha}^{\dagger} \left( -i\omega + E(\kv) \right)
    f_{k,\alpha}^{\phantom{\dagger}} +
    U \int \mathcal{D}k_1 \mathcal{D}k_2 \mathcal{D}q \;
    f_{k_1,\alpha}^{\dagger} f_{k_1+q,\alpha}^{\phantom{\dagger}}
                   f_{k_2,\beta}^{\dagger} f_{k_2-q,\beta}^{\phantom{\dagger}}
  \nonumber \ ,
\end{equation}
where $k=(\omega,\kv)$, $\mathcal{D}k = \dd \omega \dd^d \kv / (2\pi)^{d+1}$, and $E(\kv) = E_0 + \frac{k^2}{2m^*}$. Since the only fermion field $f$ in this theory lives in a vacuum state, the fermion propagator acquires no renormalization at all orders of perturbation theory (all Feynman loop diagrams are zero). Only the interaction vertex is renormalized by high energy fluctuations, from an exactly summable geometric series of ladder diagrams. This allows deriving the exact RG equations to all orders of perturbation theory \cite{SubirQPT}:
\begin{equation}
\frac{\dd E_g}{\dd l} = 2 E_g \qquad , \qquad \frac{\dd U}{\dd l} = (2-d)U - \Pi U^2 \ ,
\end{equation}
where $E_g = E_0 + U$ is the effective band-gap and $\Pi$ is a positive cutoff-dependent constant. The unitarity fixed point is $E^*_g=0$, $U^*=-\epsilon\Pi^{-1}$ in $d = 2+\epsilon \ge 2$ dimensions. Having the exact RG equations, we find the flow of couplings with the scale parameter:
\begin{equation}
  U(l) = \left\lbrace
    \begin{array}{lcl}
      \frac{U(0)}{1+\Pi U(0)l} & , & d=2 \\[0.1in]
      \frac{U(0)}{\lbrack 1+\Pi U(0) \rbrack e^l-\Pi U(0)} & , & d=3
    \end{array}
  \right\rbrace
\end{equation}

In two dimensions any initially negative $U$ flows toward $-\infty$, while in three dimensions only sufficiently large attractive $U$ undergoes a run-away flow. In both cases, however, the run-away flow escalates at a finite $l$ where $U(l)\to-\infty$ has a vertical asymptote. This indicates a breakdown of the RG. The working assumption behind the RG was that high-energy fermions are not paired. When $U\to-\infty$, even the high-energy fermions must form Cooper pairs, and this happens at a finite scale $l$. At very large attractive $U$ one must abandon the theory (\ref{CritTheory1}) and continue the RG using a theory which contains bosonic fields at all energy scales. This clearly affects the universality class of the superconducting transition, as order parameter fluctuations become qualitatively important.

The bosonic effective theory in question can be derived by rewriting the model (\ref{model1}) in the basis of Wannier states. Assuming that fermions live only at ``high energies'', either due to their band-gap or pairing, we can integrate them out and arrive at a lattice boson model:
\begin{equation}\label{EffModel}
S_{\Tr{eff}}^{(2)} = \int \dd\tau \biggl\lbrack \sum_{ij} b_i^{\dagger}
    K_{ij}\left(\frac{\partial}{\partial\tau}\right) b_j^{\phantom{\dagger}}
  + \sum_{ijkl} \mathcal{U}_{ij}^{kl} b_i^{\dagger} b_j^{\dagger}
    b_k^{\phantom{\dagger}} b_l^{\phantom{\dagger}} + \cdots \biggr\rbrack \ .
\end{equation}
Here $K_{ij}(x)$ is a function analytic at $x=0$, and the bosonic fields $b_i$ live on lattice sites and have multiple orbital states, both labeled by indices $i,j$. On the other side of the superfluid transition is a bosonic Mott insulator with gapped but long-lived bosonic excitations.

This theory is valid up to the energy scales at which the boson excitations cannot decay into free fermion pairs. A bosonic insulator can be accommodated as long as the gap for collective excitations $\Delta_b$ lies below the threshold for two fermion creation $2\Delta_f$. This threshold can be estimated in the mean-field theory from the frequency dependence of the bosonic Green's function $G(\qv,\Omega)=-\Pi^{-1}(\qv,\Omega)$. The Fig.\ref{PiPlot}(b) illustrates a Bose-insulator regime in two dimensions at $T=0$ in which the lowest excitations are infinite-lifetime gapped bosonic modes (darker shades correspond to larger positive $2\Delta_f-\Delta_b$). The superconducting transition occurring from this bosonic insulator is fluctuation-driven. The transition is effectively in the BEC regime, even though the required attractive interaction $U$ can be very small.

The most familiar Mott insulators have an integer number of bosons localized at each site due to strong repulsive forces among them \cite{Fisher1989a}. However, Mott insulators can be stabilized at other commensurate boson densities leading to translational symmetry breaking \cite{balents05}. In the present model the density of bosons is an integer per site since they originate from pairing of fermions in a fully populated band. However, the model is inescapably a multi-orbital one, and there is no constraint on the density of bosons in any particular orbital. The multi-orbital character of the effective theory (\ref{EffModel}) is tied to the same properties of the vertex functions (\ref{Ver}) which led to the PDW pairing instability. Therefore, we conjecture that the microscopic models such as (\ref{model1}) and the corresponding effective theories (\ref{EffModel}) can contain stable bosonic Mott insulating phases, adjacent to the PDW superfluid, which break translational symmetry with density and orbital orders.

\section{Acknowledgments}

We acknowledge support from NIST (PN, grant 70NANB7H6138, Am 001), NSERC of Canada (AAB and AP) and the
Sloan Foundation (AP).

\end{document}